\documentclass[10pt,twocolumn,aps,pra,groupedaddress,superscriptaddress,amsmath,amssymb,showpacs,english]{revtex4-2}
\usepackage{graphicx}
\usepackage{dcolumn}
\usepackage{bm}
\usepackage[T1]{fontenc}
\usepackage{textcomp}
\usepackage{mathrsfs}
\usepackage{amsmath}
\usepackage{bbm}
\usepackage[normalem]{ulem}
\usepackage{amsbsy}
\usepackage{braket}
\usepackage{amstext}
\usepackage{amssymb}
\usepackage{setspace}
\usepackage{esint}
\usepackage{xcolor,colortbl}
\usepackage[colorlinks,citecolor=blue,urlcolor=blue,linkcolor=blue,hypertexnames=true]{hyperref} 
\usepackage{babel}
\usepackage{booktabs}
\usepackage{tikz}
\usepackage{mciteplus}
\begin{document}

\title{Entropic uncertainty under indefinite causal order and input-output direction}

\author{G\"{o}ktu\u{g} Karpat}
\email{goktug.karpat@sabanciuniv.edu}
\affiliation{Faculty of Engineering and Natural Sciences, Sabanci University, Tuzla, Istanbul 34956, Turkey}

\date{\today}

\begin{abstract}

Entropic uncertainty relations quantify the limits on the predictability of quantum measurements. When the measured system is correlated with a quantum memory, these limits are described by the memory-assisted entropic uncertainty relation (MA-EUR). We examine the behavior of MA-EUR when the memory qubit undergoes noisy dynamics implemented via high-order controlled processes, namely, the quantum switch and the quantum time-flip. We consider a setting in which the control qubit is the very system on which the measurements are performed, while the target qubit serves as a noisy quantum memory. Focusing on Pauli channels, we show that feeding them into the quantum switch and the quantum time-flip can significantly reduce the total entropic uncertainty as compared to their direct application. Our results reveal that indefinite causal order and input-output direction can serve as resources to mitigate the effects of noise in the context of MA-EUR and its applications.

\end{abstract}

\maketitle

\section{Introduction}

Uncertainty relations are an inherent aspect of the formulation of quantum theory. It was Heisenberg who first discussed the notion of uncertainty for measurements of non-commuting observables such as position and momentum in 1927~\cite{heisenberg27}. Soon after, Kennard mathematically formalized an uncertainty relation for the position and momentum observables, i.e. $(\Delta x)^2 (\Delta p)^2 \geq \hbar^2/4$, in terms of the variances of the outcomes of their measurements~\cite{kennard27}. It was later on generalized by Robertson for arbitrary observables $Q$ and $R$ as $(\Delta Q)^2 (\Delta R)^2 \geq \frac{1}{4} \left| \langle [Q, R] \rangle \right|^2$, which is commonly accepted in textbooks as the standard form of the uncertainty relation~\cite{robertson29}. A year later, Schrödinger strengthened the lower bound by adding a new term on the right-hand side involving the anticommutator of the considered observables~\cite{schrodinger30}. The uncertainty bounds appearing in these relations naturally depend on the state of quantum system on which measurements are performed. Thus, they become trivial in case the quantum state of the system is an eigenstate of either $Q$ or $R$, even when $Q$ and $R$ are incompatible. Many years later, motivated by this drawback, Maccone and Pati proposed an uncertainty relation, in terms of the sum of the variances of the observables $Q$ and $R$  rather than their products, which in general has a non-trivial uncertainty bound even when one of the observables has a definite value for the measured state~\cite{maccone14}. All the same, the bounds in variance-based relations are inevitably state dependent and only quantify the spread of the results of the measurements.

Within the framework of information theory, entropy provides a fundamental and operationally relevant quantifier of uncertainty rather than variance. Whereas variance measures the spread of possible measurement outcomes, entropy characterizes the intrinsic unpredictability of those outcomes. This perspective led to the formulation of entropic uncertainty relations (EURs)~\cite{hirschman57,beckner75,białynicki-birula75,wehner10}, in which the total uncertainty associated with two incompatible observables is expressed in terms of the Shannon entropies of their respective measurement outcome distributions. The most well-known EUR was first proposed by Deutsch~\cite{deutsch83}, and following a conjecture by Kraus~\cite{kraus87}, was later refined by Maassen and Uffink~\cite{maassen88}, yielding
\begin{equation} \label{MU-EUR}
H(Q) + H(R) \geq -\log_2{c(Q,R)},
\end{equation}
where $H(O) = - \sum_k p_k(O) \log_2 p_k(O)$ denotes the Shannon entropy of the considered observables  $O\in \{Q,R\}$, and $p_k(O) = \mathrm{Tr}[\varrho\, \Pi_k^{O}]$ with $\Pi_k^{O} = |\psi_k^{O}\rangle \langle \psi_k^{O}|$ giving the probability of obtaining the $k$th outcome when the observable $O$ is measured on the density operator $\varrho$ of the system. The projection operator $\Pi_k^{O}$ is constructed from the eigenstates of the observable $O$. Complementarity of the observables $Q$ and $R$ is $c(Q,R)=\max_{i,j}|\langle \psi_i^Q|  \psi_j^R \rangle|^2$, where $|\psi_i^Q\rangle$ and $|\psi_j^R\rangle$ are their eigenstates. As the complementarity only depends on the observables $Q$ and $R$, it is clear that the lower bound here is state independent. 

An intriguing extension of the EUR in Eq.~\eqref{MU-EUR} has been introduced by Berta et al.~\cite{berta10}, where the particle $A$ being measured, is considered to be entangled with a quantum memory $B$. In this setting, the memory-assisted entropic uncertainty relation (MA-EUR) takes the form
\begin{equation} \label{MA-EUR}
S(Q|B) + S(R|B) \geq -\log_2 c(Q,R) + S(A|B),
\end{equation}
where $S(\varrho)=-\operatorname{Tr}[\varrho \log_2 \varrho]$ is the von Neumann entropy, and $S(A|B)=S(\varrho_{AB})-S(\varrho_{B})$ represents the conditional entropy. On the other hand, $S(O|B)=S(\varrho_{OB})-S(\varrho_{B})$, with $O\in \{Q,R\}$, denotes the conditional entropies of the post-measurement states $\varrho_{OB}=\sum_k (\Pi_k^{O}\otimes \mathbb{I}) \varrho_{AB} (\Pi_k^{O}\otimes \mathbb{I}) $, after a measurement of the observable $O$ is performed on the system $A$, where $\mathbb{I}$ is the identity operator acting on memory $B$. Notice that the lower bound Eq.~\eqref{MA-EUR} contains both state independent and state dependent terms. MA-EUR in Eq.~\eqref{MA-EUR} was experimentally verified~\cite{li11,prevedel11}. Furthermore, it was shown to be relevant for numerous subjects in  quantum information science, such as quantum key distribution~\cite{berta10,pramanik13}, witnessing~\cite{xu12,liu21} and sharing entanglement~\cite{hu23}, quantum batteries~\cite{song22, song24}, quantum teleportation~\cite{hu12,sun22}, and more~\cite{coles17,wang19,Ming20,Wu22,Wang24a,Wang24b}.

Since realistic quantum systems are prone to the detrimental effects of the environment~\cite{breuer02}, MA-EUR has also been investigated under noisy quantum evolutions, which are typically modeled by quantum channels~\cite{nielsen11}. Given the relevance of MA-EUR in quantum information tasks, understanding the impact of environmental noise on both the total entropic uncertainty and its lower bound constitutes a problem of practical interest. In order to study this problem, different dynamical regimes such as Markovian and non-Markovian, as well as noise settings such as local and global channels, have been considered in the recent literature~\cite{xu12-2,karpat15,karpat18,haseli20,haddadi21,khedr21,ming17,yao20,chen19,basit24,hamad25}. In general, it has been observed that environmental noise tends to increase the total entropic uncertainty by diminishing quantum correlations between the measured system and the quantum memory. However, quantum channels describing the dynamics induced by certain structured or non-Markovian environments can help reduce the uncertainty~\cite{basit24,hamad25,karpat15}.

In classical physics, causal relations between events are well defined. Present events are determined by those in the past and, in turn, they influence those in the future. Still, quantum theory admits a more general structure, in which quantum operations may occur without a fixed causal order. It was argued that there exist causally non-separable processes that cannot be described by any definite ordering of events~\cite{Chiribella2013,Oreshkov2012,Brukner2014,Araujo2015,Oreshkov2016,Barrett2021}. A paradigmatic example of such a process is the quantum switch~\cite{Chiribella2013}, which coherently controls the order in which two quantum operations are applied. In essence, quantum switch places two alternative causal orders into  superposition, leading to scenarios where the sequence of quantum channels itself becomes indefinite. Quantum switch was experimentally implemented in various platforms~\cite{Procopio2015,Rubino2017,Guo2020,Cao2023,Goswami2018,Goswami2020,Tang25,Rubino2022}. Moreover, the exploitation of indefinite causal order was shown to offer advantages in several areas, such as quantum metrology~\cite{Zhao2020,Chapeau-Blondeau2021,Liu2023,Zhou2024}, channel discrimination~\cite{Chiribella2012}, communication complexity~\cite{Guerin2016}, query complexity~\cite{Colnaghi2012,Araujo2014,Renner2022}, noisy transmission of information~\cite{Ebler2018,Procopio2019,Goswami2020a,Caleffi2020,Bhattacharya2021,Chiribella2021a,Chiribella2021b,Sazim2021,Mukhopadhyay2020}, quantum thermodynamics~\cite{Felce2020,Guha2020,Simonov2022,Liu2022}, and non-Markovian processes~\cite{karpat24,Mukherjee24,Anand25}.

Besides, while classical processes evolve along a fixed direction in time, quantum theory allows for a more symmetric treatment where the inputs and outputs of a process can be interchanged~\cite{Aharonov1964,Aharonov1990,Aharonov2002,Hardy2007,Oeckl2008,Svetlichny2011,Lloyd2011,Genkina2012,Oreshkov2015,Silva2017}. Building on this idea, a formal framework was recently introduced to characterize quantum operations under time reversal~\cite{Chiribella2022}. This framework makes it possible to coherently superpose a forward process with its corresponding backward (input-output inverted) counterpart, resulting in a quantum evolution having an indefinite input-output direction. A prototypical example of such an operation is the quantum time-flip~\cite{Chiribella2022}, which applies a quantum channel to a system in a superposition of its forward and backward modes. Quantum time-flip was experimentally realized in photonic setups, where it demonstrated better performance in discrimination games over any other strategy restricted to a definite time direction~\cite{Stromberg2024,Guo2024}. Moreover, it was shown to offer significant advantages for communication through noisy channels~\cite{Liu2023b} and quantum metrology~\cite{Agrawal2025}.

In this work, we study the behavior of the entropic uncertainty assuming that the memory qubit $B$ is subjected to a noisy channel implemented through either the quantum switch or the quantum time-flip superchannels. In the standard formulations of these processes, the control qubit $A$ is regarded as an ancillary. After coherently superposing alternative causal orders or input-output directions, the control qubit $A$ is typically first measured and then traced out in order to investigate the effective dynamics that acts on the target system $B$. In our setting, however, the control qubit $A$ is not an ancillary but instead it represents the very system on which a pair of measurements will be performed in the context of MA-EUR, while the target qubit $B$ assumes the role of a noisy quantum memory. Even though our framework remains consistent with the standard descriptions of higher-order quantum channels such as the quantum switch and the quantum time-flip, it actually corresponds to a distinct operational scenario, that is, rather than characterizing the effective channel on the target system $B$ alone, we explore how indefinite causal order and indefinite input-output direction influence the total entropic uncertainty in the MA-EUR when a pair of measurements are non-selectively performed on the control qubit $A$. We initially focus on the class of Pauli channels directly acting on the qubit $B$. Then, by feeding these channels into the quantum switch and the quantum time-flip, we show that both of these processes can significantly reduce the total entropic uncertainty, in comparison with the case in which the Pauli channel directly affects the memory qubit.

This paper is organized as follows. In Sec.~\ref{sec2}, we describe the Pauli channels. Sec.~\ref{sec3} defines the quantum switch and quantum time-flip. In Sec.~\ref{sec4}, we present our main results. Sec.~\ref{sec5} serves as a summary of our results.

\section{Pauli Channels} \label{sec2}

A quantum channel is a completely positive and trace-preserving (CPTP) linear map defined from the input Hilbert space $\mathcal{H}_{\mathrm{in}}$ to the output Hilbert space $\mathcal{H}_{\mathrm{out}}$ such that $\mathcal{E}:\mathcal{B}(\mathcal{H}_{\mathrm{in}}) \rightarrow \mathcal{B}(\mathcal{H}_{\mathrm{out}})$, where $\mathcal{B}(\mathcal{H}_{\mathrm{in}})$ and $\mathcal{B}(\mathcal{H}_{\mathrm{out}})$ denote the set of linear operators on these Hilbert spaces. Every channel admits a Kraus decomposition of the form
\begin{equation}
\mathcal{E}(\varrho) = \sum_{i} K_i \varrho K_i^\dagger,
\qquad \varrho \in \mathcal{D}(\mathcal{H}_{\mathrm{in}}),
\end{equation}
where the operators $K_i \in \mathcal{B}(\mathcal{H}_{\mathrm{in}},\mathcal{H}_{\mathrm{out}})$ are called Kraus operators that are linear maps from $\mathcal{H}_{\mathrm{in}}$ to $\mathcal{H}_{\mathrm{out}}$~\cite{Kraus1971,Kraus1983}. Also, $\mathcal{D}(\mathcal{H}_{\mathrm{in}})$ stands for the set of density operators on $\mathcal{H}_{\mathrm{in}}$. The trace preservation property requires that
\begin{equation}
\sum_i K_i^\dagger K_i = \mathbb{I}_{\mathrm{in}}, 
\end{equation}
where $\mathbb{I}_{\mathrm{in}}$ is the identity operator on $\mathcal{H}_{\mathrm{in}}$. In the special case, $\mathcal{H}_{\mathrm{in}}=\mathcal{H}_{\mathrm{out}}=\mathcal{H}$, the Kraus operators are simply square matrices acting on a common Hilbert space $\mathcal{H}$.

In our treatment, we focus on single-qubit Pauli channels, where the input and output Hilbert spaces coincide with $\dim\mathcal{H}=2$. These quantum channels are defined as convex combinations of the identity operation and error processes generated by the Pauli operators. Formally, a Pauli map $\Lambda:\mathcal{B}(\mathcal{H}) \rightarrow \mathcal{B}(\mathcal{H})$  is defined as
\begin{equation} \label{PauliChan}
\Lambda(\varrho) = \sum_{i} q_i\, \sigma_i \varrho \sigma_i,
\end{equation}
where $i \in \{0,x,y,z\}$, $\sigma_0=\mathbb{I}$ and $\sigma_x,\sigma_y,\sigma_z$ are respectively the standard Pauli operators in $x,y,z$ directions. The real coefficients $q_i$ form a probability distribution, which implies that each $q_i \geq 0$ and $\sum_{i=0}^3 q_i = 1$. While $q_0$ is the probability of no error, $q_x, q_y$ and $q_z$ correspond to bit-flip, bit-phase-flip, and phase-flip error probabilities, respectively. We note that Pauli channels are unital, meaning $\Lambda(\mathbb{I}) = \mathbb{I}$, since each Pauli operator is unitary and conjugation by a unitary leaves the identity operator invariant. Indeed, it is convenient to parametrize $q_i$ as
\begin{equation} \label{alphap}
q_0 = 1-p, 
\quad q_x = \alpha_x p, 
\quad q_y = \alpha_y p, 
\quad q_z = \alpha_z p,
\end{equation}
where $p \in [0,1]$ denotes the overall error probability and $\vec{\alpha}=(\alpha_x,\alpha_y,\alpha_z)$ is a bias vector with non-negative components summing to unity. With this parametrization, it is possible to interpolate between isotropic depolarizing noise ($\alpha_x=\alpha_y=\alpha_z=1/3$) and anisotropic noise that is biased toward specific types of error.

Any qubit state can be expressed in Bloch form,
\begin{equation}
\varrho = \tfrac12\!\left(\mathbb{I} + r_x \sigma_x + r_y \sigma_y + r_z \sigma_z\right),
\end{equation}
where $\vec{r}=(r_x,r_y,r_z)$ is called the Bloch vector. Under the Pauli channel given in Eq.~\eqref{PauliChan}, the output state is
\begin{equation}
\Lambda(\varrho) = \tfrac12\!\left(\mathbb{I} + \lambda_x r_x \sigma_x + \lambda_y r_y \sigma_y + \lambda_z r_z \sigma_z\right),
\label{eq:bloch-action}
\end{equation}
since $\Lambda[\sigma_i]=\lambda_i\sigma_i$ with $\lambda_0=1$. The shrinking parameters for Bloch vector entries $(r_x,r_y,r_z)$ are given as~\cite{Siudzinska19}
\begin{align} \label{lambdas}
\lambda_x &= q_0+q_x-q_y-q_z = 1-2(1-\alpha_x)p, \nonumber \\
\lambda_y &= q_0-q_x+q_y-q_z = 1-2(1-\alpha_y)p, \nonumber \\
\lambda_z &= q_0-q_x-q_y+q_z = 1-2(1-\alpha_z)p,
\end{align}
using the parametrization introduced in Eq.~\eqref{alphap}. Hence, the action of the Pauli channel is an anisotropic contraction of the Bloch ball into an axis-aligned ellipsoid. The parameters $(\lambda_x,\lambda_y,\lambda_z)$ must satisfy certain inequalities so that the map $\Lambda$ defines a valid quantum channel. For Pauli type maps,
complete positivity is equivalent to the satisfaction of the Fujiwara--Algoet inequalities~\cite{Fujiwara99}
\begin{equation}
|1 \pm \lambda_z| \,\geq\, |\lambda_x \pm \lambda_y|,
\label{eq:FA}
\end{equation}
which guarantees that the Bloch ball is transformed into an ellipsoid contained within the Bloch ball of valid density operators. In fact, in the parametrization in Eq.~\eqref{alphap}, the real coefficients $q_i$ are non-negative and their sum is normalized to unity, which means that the Pauli map $\Lambda$ is composed of a convex mixture of Pauli unitaries. Consequently, the map $\Lambda$ is by construction CPTP and the inequalities in Eq.~\eqref{eq:FA} are automatically satisfied.

Our focus on the Pauli channels in this work is motivated by three main considerations. First, they constitute a canonical noise model in quantum information, encapsulating various fundamental error mechanisms ranging from biased bit-flips and phase-flips to the isotropic depolarizing channel. Second, their defining property of mapping Pauli operators to Pauli operators makes the resulting dynamics analytically tractable and allows closed-form evaluation of the conditional entropies which appear in the MA-EUR. Finally, the formulation of the quantum time-flip is naturally tailored to bidirectional processes and thus necessitates unitality, a requirement that is automatically satisfied by Pauli channels.

\section{Quantum Switch and Time-Flip} \label{sec3}

This section introduces processes in which causal order and input-output direction are no longer fixed. Such exotic processes are formalized through higher-order transformations also known as superchannels that operate on quantum channels themselves, rather than acting directly on density operators. As the quantum switch implements indefinite causal order by placing two operations in a superposition of their orders, the quantum time-flip operation yields indefinite input–output direction via the superposition of forward and reverse processes.

\subsection{Quantum Switch} \label{switch-sec}

Let $\Lambda_1$ and $\Lambda_2$ be two qubit channels having two sets of Kraus operators $\{M^{(1)}_i\}$ and $\{M^{(2)}_j\}$, respectively. Quantum switch acts on a bipartite state described by a density operator $\varrho_{AB}$, where the first qubit ($A$) serves as a control, determining the ordering of the sequence of the channels $\Lambda_1$ and $\Lambda_2$ acting on the second qubit ($B$). With these considerations, the switched channel can be constructed using the set of Kraus operators as~\cite{Chiribella2013}
\begin{equation} \label{kraussw}
S_{ij}
= |0\rangle\langle 0|_A \otimes M^{(2)}_i M^{(1)}_j
 + |1\rangle\langle 1|_A \otimes M^{(1)}_j M^{(2)}_i,
\end{equation}
which define the bipartite operation given by
\begin{equation} \label{sschan}
\mathcal{S}(\varrho_{AB})=\sum_{i,j} S_{ij}\,\varrho_{AB}\,S_{ij}^\dagger. 
\end{equation}
Here, the control basis is formed by the two eigenstates $\{|0\rangle,|1\rangle\}$ of the Pauli matrix $\sigma_z$. In the context of quantum switch, it is convenient to express a given bipartite density operator $\varrho_{AB}$ as a $2 \times 2$ block matrix in the basis of the control qubit $A$ as
\begin{equation} \label{block-mat}
\varrho_{AB}=\sum_{a,b} |a\rangle\langle b|_{A} \otimes X_{ab},
\quad
X_{ab} = \langle a|\varrho_{AB}|b\rangle_{A},
\end{equation}
where $a,b\in\{0,1\}$ and $X_{ab}$ is an operator (a $2\times 2$ matrix) on the target qubit $B$. Then, the action of the bipartite channel $\mathcal{S}$ on the density operator $\varrho_{AB}$ is described as
\begin{align} 
\mathcal{S}(\varrho_{AB}) ={}&|0\rangle\langle 0|_A \otimes \mathcal{S}_{00}(X_{00}) + |0\rangle\langle 1|_A \otimes \mathcal{S}_{01}(X_{01}) \nonumber \\
& + |1\rangle\langle 0|_A \otimes \mathcal{S}_{10}(X_{10}) + |1\rangle\langle 1|_A \otimes \mathcal{S}_{11}(X_{11}), \nonumber 
\end{align}
or equivalently in the block matrix form,
\begin{equation} \label{sup-mat}
\mathcal{S}(\varrho_{AB})=
\begin{pmatrix} 
\mathcal{S}_{00}(X_{00}) & \mathcal{S}_{01}(X_{01})\\
\mathcal{S}_{10}(X_{10}) & \mathcal{S}_{11}(X_{11})
\end{pmatrix}_{\!A},
\end{equation}
where the block superoperators read
\begin{align} \label{sup-op-switch}
\mathcal S_{00}(X_{00}) &= \sum_{i,j} M^{(2)}_i M^{(1)}_j\,X_{00}\,M^{(1)\dagger}_j M^{(2)\dagger}_i, \nonumber \\ 
\mathcal S_{11}(X_{11}) &= \sum_{i,j} M^{(1)}_j M^{(2)}_i\,X_{11}\,M^{(2)\dagger}_i M^{(1)\dagger}_j,\nonumber  \\ 
\mathcal S_{01}(X_{01}) &= \sum_{i,j} M^{(2)}_i M^{(1)}_j\,X_{01}\,M^{(2)\dagger}_i M^{(1)\dagger}_j, \nonumber \\ 
\mathcal S_{10}(X_{10}) &= \sum_{i,j} M^{(1)}_j M^{(2)}_i\,X_{10}\,M^{(1)\dagger}_j M^{(2)\dagger}_i.
\end{align}
\vspace{-0.05cm}
Note that here we do not assume that the control qubit $A$ and the target qubit $B$ are in a factorized state, differently from what has been usually done in the literature. 

Using the block matrix representation, we can see how the measurement of the control qubit $A$ will determine the evolution of the target qubit $B$ even if their composite state is entangled. Let us first suppose that the control qubit $A$ is measured in $\sigma_z$ basis, which projects its state onto one of the two possible eigenstates $\{|0\rangle,|1\rangle\}$. If the state is projected to the state $|0\rangle$ after the measurement, then the state of the target qubit becomes
\begin{equation}
\varrho_{B}^{(0)} = \frac{\mathcal{S}_{00}(X_{00})}{p_0}, \quad p_0=\operatorname{Tr}[\mathcal{S}_{00}(X_{00})]=\operatorname{Tr}[X_{00}],
\end{equation}
where $p_0$ is the probability of obtaining this outcome. In this case, Eqs.~\eqref{sup-op-switch} shows that the resulting state of the target $B$ is given by applying the channel $\Lambda_2 \circ \Lambda_1$ to the initial conditional state $X_{00}$, which represents the state of the target $B$, given that the control $A$ is found in the state $|0\rangle$. On the other hand, if the initial state is factorized, that is, $\varrho_{AB}=\varrho_{A} \otimes \varrho_{B}$, then $X_{00}=\langle 0|\varrho_{A}|0\rangle\varrho_B$, which is $\varrho_B$ scaled by the probability of the control being in $|0\rangle$. Hence, for factorized inputs, $\mathcal{S}_{00}(\varrho_B) = (\Lambda_2 \circ \Lambda_1)(\varrho_B)$, meaning that first $\Lambda_1$ and then $\Lambda_2$ act on the state of the target qubit $\varrho_B$ in a definite order. Considering the case where the state of the control $A$ is projected to the state $|1\rangle$ after the measurement is performed, we similarly find that the state of the target qubit becomes
\begin{equation}
\varrho_{B}^{(1)} = \frac{\mathcal{S}_{11}(X_{11})}{p_1}, \quad p_1=\operatorname{Tr}[\mathcal{S}_{11}(X_{11})]=\operatorname{Tr}[X_{11}],
\end{equation}
where $p_1$ is the probability of this outcome. It can be seen from Eqs.~\eqref{sup-op-switch} that final state of the target $B$ is the result of applying the channel $\Lambda_1 \circ \Lambda_2$ to the initial conditional state $X_{11}$. For a factorized input state, this simplifies to the usual definite causal order $\mathcal{S}_{11}(\varrho_B) = (\Lambda_1 \circ \Lambda_2)(\varrho_B)$, where first the channel $\Lambda_2$ and then $\Lambda_1$ act on the target state $\varrho_B$. Therefore, in cases where the measurement is performed in $\sigma_z$ basis, only the diagonal components of Eq.~\eqref{sup-mat}, namely, the superoperators $\mathcal S_{00}$ and $\mathcal S_{11}$ play a role, and the novel features that have their roots in the coherent superpositions of $\Lambda_1$ and $\Lambda_2$ cannot be observed. Let us next suppose that the qubit $A$ is measured in $\sigma_x$ basis, which projects its state onto one of the two possible eigenstates $\{|+\rangle,|-\rangle\}$, where $|\pm\rangle = (1/\sqrt{2})(|0\rangle \pm |1\rangle)$. Depending on the outcome, the (non-normalized) state of the target qubit $B$ takes one of the following two forms:
\begin{align} \label{mynote}
\varrho_B^{(+)} &= \tfrac{1}{2}\Big[
\mathcal{S}_{00}(X_{00}) + \mathcal{S}_{11}(X_{11})
+ \mathcal{S}_{01}(X_{01}) + \mathcal{S}_{10}(X_{10})
\Big], \nonumber\\[4pt]
\varrho_B^{(-)} &= \tfrac{1}{2}\Big[
\mathcal{S}_{00}(X_{00}) + \mathcal{S}_{11}(X_{11})
- \mathcal{S}_{01}(X_{01}) - \mathcal{S}_{10}(X_{10})
\Big]. \raisetag{-5pt}
\end{align}
As can be easily seen, the state of the target $B$ is now a coherent sum of all four block components from Eq.~\eqref{sup-mat}. It not only depends on the outputs of the two definite order paths $\mathcal{S}_{00}$ and $\mathcal{S}_{11}$ but also on the off-diagonal terms $\mathcal{S}_{01}$ and $\mathcal{S}_{10}$ that represent the interference between them. This interference between the two causal orders $(\Lambda_2 \circ \Lambda_1)$ and $(\Lambda_1 \circ \Lambda_2)$ is the signature of a quantum process with no definite causal order. Lastly, if the control qubit $A$ is discarded without being observed, i.e. traced out, then it is clear from Eq.~\eqref{sup-mat} that the state of the target $B$ is given by an incoherent mixture of the two definite causal orders as $\varrho_B = \operatorname{Tr}_A[\mathcal{S}(\varrho_{AB})] = \mathcal{S}_{00}(X_{00}) + \mathcal{S}_{11}(X_{11})$.  In summary, the switch uses a control qubit to dictate the causal order of operations, and a measurement on this control either collapses the process into a definite sequence of operations or reveal the interference generated by the superposition of both possible orders. Note that the dynamics generated by the quantum switch is independent of the Kraus representation of the input channels, and it gives rise to novel evolutions that are inaccessible in frameworks with a definite causal structure~\cite{Chiribella2013}. 

\subsection{Quantum Time-Flip}

The second superchannel that we consider in our work is known as the quantum time-flip, which implements a quantum process with indefinite input-output direction by superposing a certain forward channel with the corresponding backward channel. It is possible to think of some quantum processes as being physically reversible. For instance, polarization of a photon might be rotated by passing through a crystal, and it is natural to consider the photon being able to pass through the crystal in the opposite direction. Such reversible processes are formalized as bidirectional processes~\cite{Chiribella2022}. It has been established that a quantum channel $\Phi$ is bidirectional if a valid backward channel $\theta(\Phi)=\Phi^T$ exists, where $\theta$ represents the input-output inversion transformation. Based on a few symmetry arguments and physically motivated axioms~\cite{Chiribella2022}, it has been proved that a quantum process is bidirectional if and only if it can be described by a unital quantum channel having a Kraus representation,
\begin{equation}
\Phi(\varrho)=\sum_i M_i \varrho M_i^\dagger,
\end{equation}
where, in addition to the condition $\sum_i M_i^\dagger M_i= \mathbb{I}$, which ensures trace preservation, $\sum_i  M_i M_i^\dagger= \mathbb{I}$ is also satisfied due to unitality. For a given bidirectional channel $\Phi$, the quantum time-flip is constructed as
\begin{equation} \label{tfchann}
\mathcal{F}(\varrho_{AB})=\sum_{i} F_{i}\,\varrho_{AB}\,F_{i}^\dagger. 
\end{equation}
with the Kraus operators
\begin{equation}
F_i=|0\rangle\langle 0|_{A}\otimes M_i + | 1\rangle\langle 1|_{A}\otimes \theta(M_i),
\end{equation}
which acts on a bipartite quantum system where the first qubit $A$ plays the role of a control and the second qubit $B$ that of a target, similarly to the case of quantum switch. It has also been demonstrated that the input-output inversion operation $\theta$ for quantum channels must be represented by the transposition map~\cite{Chiribella2022}, i.e. $\theta(M_i)=M_i^T$, which guarantees that the map $\mathcal{F}$ is independent of the Kraus representation and ensures that it is a valid CPTP map. Thus, the input-output inverted channel reads
\begin{equation}
\Phi^T(\varrho)=\sum_i M_i^{T}\varrho (M_i^{T})^{\dagger}=\sum_i M_i^{T}\varrho M_i^{*},
\end{equation}
and the quantum time-flip Kraus operators take the form
\begin{equation}
F_i=|0\rangle\langle 0|_{A}\otimes M_i + | 1\rangle\langle 1|_{A}\otimes M_i^T.
\end{equation}
Using the same block matrix representation introduced in Eq.~\eqref{block-mat}, the action of the bipartite channel $\mathcal{F}$ on the density operator $\varrho_{AB}$ can be expressed as
\begin{align} 
\mathcal{F}(\varrho_{AB}) ={}& |0\rangle\langle 0|_A \otimes \mathcal{F}_{00}(X_{00}) + |0\rangle\langle 1|_A \otimes \mathcal{F}_{01}(X_{01}) \nonumber \\
& + |1\rangle\langle 0|_A \otimes \mathcal{F}_{10}(X_{10}) + |1\rangle\langle 1|_A \otimes \mathcal{F}_{11}(X_{11}), \nonumber 
\end{align}
or equivalently in the block matrix form,
\begin{equation}
\mathcal{F}(\varrho_{AB})=
\begin{pmatrix} \label{sup-mat2}
\mathcal{F}_{00}(X_{00}) & \mathcal{F}_{01}(X_{01})\\
\mathcal{F}_{10}(X_{10}) & \mathcal{F}_{11}(X_{11})
\end{pmatrix}_{A},
\end{equation}
where the block superoperators are given by
\begin{align} \label{sup-op-flip}
\mathcal{F}_{00}(X_{00})&=\sum_i M_i\,X_{00}\,M_i^{\dagger} = \Phi(X_{00}),
\nonumber \\
\mathcal{F}_{11}(X_{11})&=\sum_i M_i^{T}\,X_{11}\,M_i^{*} = \Phi^{T}(X_{11}),
\nonumber \\
\mathcal{F}_{01}(X_{01})&=\sum_i M_i\,X_{01}\,M_i^{*},
\nonumber \\
\mathcal{F}_{10}(X_{10})&=\sum_i M_i^{T}\,X_{10}\,M_i^{\dagger}.
\end{align}
Similarly to the case of the quantum switch, the block-matrix representation in Eq.~\eqref{block-mat} allows us to analyze the role of the control qubit in the time-flip scenario. In particular, measurement of the control qubit $A$ in different bases, or its complete discarding, leads to dynamics of the target qubit $B$ that closely resembles the structure encountered in Sec.~\ref{switch-sec}. The only difference is that the block superoperators are now those given in Eqs.~\eqref{sup-op-flip}, corresponding to forward and backward channels rather than two different orders of sequential channels. For example, measurement in the $\sigma_z$ basis shows that provided the control qubit $A$ is projected onto $|0\rangle$ ($|1\rangle$), the forward channel $\Phi$ (backward channel $\Phi^T$), is applied to the corresponding conditional input $X_{00}$ ($X_{11}$), so that the diagonal blocks $\mathcal{F}_{00}$ and $\mathcal{F}_{11}$ encode the two definite directions of the process. On the other hand, measurement of the qubit $A$ in the $\{|+\rangle,|-\rangle\}$ basis leads to an output state that involves both diagonal and off-diagonal block superoperators, $\mathcal{F}_{01}$ and $\mathcal{F}_{10}$, and thus capturing interference between the forward and backward evolutions in direct analogy with the role of $\mathcal{S}_{01}$ and $\mathcal{S}_{10}$ in the switch. Finally, if $A$ is discarded without measurement, we end up with an incoherent mixture of the two definite directions, $\varrho_B = \operatorname{Tr}_A[\mathcal{F}(\varrho_{AB})] = \mathcal{F}_{00}(X_{00}) + \mathcal{F}_{11}(X_{11})$, acting on the conditional inputs $X_{00}$ and $X_{11}$. To sum up, the time-flip uses a control to place a channel in a superposition of its forward and backward implementations. A suitable measurement on this control can either imply a definite direction or reveal the interference between them.

\section{Main Results} \label{sec4}

In this section, we present our main findings regarding the behavior of MA-EUR in Eq.~\eqref{MA-EUR} when the memory qubit $B$ is undergoing noisy dynamics. Before introducing the noise on the memory, let us first elaborate on the operational meaning of the MA-EUR as an uncertainty game between Alice and Bob, who respectively possess the particle $A$ and the quantum memory $B$, thereby providing an interpretation of the entropic bound. In this game, Alice randomly chooses to measure either $Q$ or $R$ on her particle $A$, which is entangled with Bob’s quantum memory $B$. Alice then announces which observable was measured but not the outcome of the measurement. Bob next performs a measurement on his memory qubit $B$ to infer Alice’s measurement result. Bob's uncertainty (given access to quantum memory $B$) is quantified by the conditional entropies $S(Q|B)$ and $S(R|B)$, whose sum is constrained by the lower bound in Eq.~\eqref{MA-EUR}. In absence of quantum correlations between $A$ and $B$, the conditional entropy $S(A|B)$ becomes non-negative. Thus, despite the lower bound actually still becomes tighter than the bound in relation in Eq.~\eqref{MU-EUR}, the specific quantum advantage of reducing the total uncertainty enabled by entanglement is lost.~\cite{berta10}. Interestingly, for maximally entangled states and complementary observables, both the conditional entropies and the bound vanish, so Bob can predict outcomes of Alice's measurements with certainty, which happens thanks to the fact that the conditional entropy $S(A|B)$ can take negative values for entangled states~\cite{devetak05}.

We are now ready to study how the MA-EUR in Eq.~\eqref{MA-EUR} behaves in three distinct cases: i) when the memory qubit $B$ is evolving under the effect of a Pauli channel, ii) when two identical Pauli channels are fed into the quantum switch such that we have self-switched Pauli channels on the memory $B$, and iii) when a Pauli channel is fed into quantum time-flip such that the memory qubit $B$ experiences dynamics having indefinite input-output direction.

\subsection{Single-use Pauli Channels}

Let us commence our investigation considering the case where the memory qubit $B$ evolves under a Pauli channel. For concreteness, suppose that the bipartite system $AB$ is initially in one of the Bell states, that is, $\varrho_{AB}=|\psi\rangle \langle \psi|$, where $|\psi\rangle=(1/\sqrt{2})(|00\rangle+|11\rangle)$. In the Fano-Bloch representation, this Bell state is written as
\begin{equation} \label{Bellstate}
\varrho_{AB} = \tfrac14 \left(\mathbb{I}\otimes  \mathbb{I} + \sigma_x\otimes \sigma_x - \sigma_y\otimes \sigma_y + \sigma_z\otimes \sigma_z\right).
\end{equation}
The density operator of the bipartite system $AB$, under noisy dynamics of the memory qubit $B$, is then given by $\varrho_{AB}^{\text{su}}=(\mathbb{I}_A \otimes \Lambda_B)\varrho_{AB}$, where the superscript su denotes single-use. First, note that the action of this channel on each one of the four terms, involving Pauli and identity operators, in Eq.~\eqref{Bellstate} can be calculated as
\begin{equation}
(\mathbb{I}_A \otimes \Lambda_B)(\sigma_\mu \otimes \sigma_\mu)=\sigma_\mu \otimes \Lambda(\sigma_\mu)=\lambda_\mu(\sigma_\mu \otimes \sigma_\mu).
\end{equation}
Here, the action of the Pauli channel on $\sigma_\mu$ is found from
\begin{equation}
\Lambda(\sigma_\mu)=\sum_i q_i(\sigma_i \sigma_\mu \sigma_i)=\sum_i q_i s_\mu(i)\sigma_\mu=\lambda_\mu\sigma_\mu,
\end{equation} 
where $s_\mu(i)=2(\delta_{i,0}+\delta_{i,\mu})-1$, with $i \in \{0,x,y,z\}$ and $\mu \in \{x,y,z\}$, which follows from $\{\sigma_\mu,\sigma_\nu\}=2\delta_{\mu,\nu}\mathbb{I}$, and $\lambda_\mu$ are explicitly given in Eq.~\eqref{lambdas}. As a consequence, the density operator of the system $AB$ takes the form
\begin{align} \label{evolvedrhosu}
\varrho_{AB}^{\text{su}} ={}& \tfrac14 [\mathbb{I} \otimes \mathbb{I} + \lambda_x(\sigma_x \otimes \sigma_x) \nonumber \\
& - \lambda_y(\sigma_y \otimes \sigma_y) + \lambda_z(\sigma_z \otimes \sigma_z)].  \raisetag{-5pt}
\end{align}
Therefore, the effect of the channel is to damp the terms in Eq.~\eqref{Bellstate} by the corresponding factors $\lambda_\mu$.

Having obtained the evolution of the bipartite system $AB$ when the memory $B$ is under the action of a Pauli channel, we are in a position to investigate the behavior of the MA-EUR in this setting. Throughout this work, we take the observables in Eq.~\eqref{MA-EUR} as $Q=\sigma_x$ and $R=\sigma_z$. The post-measurement state of $AB$ after a non-selective measurement in $\sigma_x$ basis on the qubit $A$ becomes
\begin{align}
\varrho_{XB}^{\text{su}} &= \sum_{k=\pm} (\Pi^{\sigma_x}_k \otimes \mathbb{I})\,\varrho_{AB}^{\text{su}}\,(\Pi^{\sigma_x}_k \otimes \mathbb{I}), \nonumber \\
& = \tfrac14 \left[\mathbb{I}\otimes  \mathbb{I} + \lambda_x(\sigma_x\otimes \sigma_x)\right],
\end{align}
where the projection operators read $\Pi^{\sigma_x}_{\pm}=(\mathbb{I}\pm\sigma_x)/2$. Similarly, a non-selective measurement in $\sigma_z$ basis on the control qubit $A$ transforms the state into
\begin{align}
\varrho_{ZB}^{\text{su}} &= \sum_{k=\pm} (\Pi^{\sigma_z}_k \otimes \mathbb{I})\,\varrho_{AB}^{\text{su}}\,(\Pi^{\sigma_z}_k \otimes \mathbb{I}), \nonumber \\
& = \tfrac14 \left[\mathbb{I}\otimes  \mathbb{I} + \lambda_z(\sigma_z\otimes \sigma_z)\right],
\end{align}
where the projection operators are $\Pi^{\sigma_z}_{\pm}=(\mathbb{I}\pm\sigma_z)/2$. Note that the density operators of the qubit $B$ before and after the application of the Pauli channel are the same, $\varrho_B=\varrho_B^{\text{su}}=\mathbb{I}/2$, due to the fact that the density operator of the bipartite system $AB$ preserves Bell-diagonal form, i.e., it remains diagonal in the Bell state basis. Therefore, the conditional entropies in Eq.~\eqref{MA-EUR} become
\begin{align}
S(X|B)_{\text{su}}=S(\varrho_{XB}^{\text{su}})-S(\varrho_B^{\text{su}}) = H_\text{bin}\Big( \frac{1+\lambda_x}{2}\Big), \\
S(Z|B)_{\text{su}}=S(\varrho_{ZB}^{\text{su}})-S(\varrho_B^{\text{su}}) = H_\text{bin}\Big( \frac{1+\lambda_z}{2}\Big),
\end{align}
where $H_\text{bin}(x)=-x \log_2 x- (1-x)  \log_2 (1-x)$ is the binary entropy. As a result, the total entropic uncertainty on the left-hand side of the MA-EUR in Eq.~\eqref{MA-EUR} is 
\begin{align} \label{Usu}
U_{\text{su}}&=S(X|B)_{\text{su}}+S(Z|B)_{\text{su}}. \nonumber \\
& = H_\text{bin}\Big( \frac{1+\lambda_x}{2}\Big) +  H_\text{bin}\Big( \frac{1+\lambda_z}{2}\Big)
\end{align}
The lower bound to this quantity can be readily obtained as the observables to be measured, namely $\sigma_x$ and $\sigma_z$, are complementary, which implies that $-\log_2C(\sigma_x,\sigma_z)=1$. Besides, the conditional entropy $S(A|B)$ is simply given by $S(A|B)=S(\varrho_{AB}^{\text{su}})-1$ since the reduced quantum state of $B$ is maximally mixed. Hence, the uncertainty bound appearing on the right-hand side of Eq.~\eqref{MA-EUR} reduces to
\begin{equation}\label{BSu}
B_{\text{su}}=S(\varrho_{AB}^{\text{su}})=-\sum\nolimits_k p_k \log_2 p_k,
\end{equation}
where $p_k$ are the four eigenvalues of the bipartite density operator $\varrho_{AB}^{\text{su}}$, which are given by
\begin{align} \label{eigvalues}
p_{1,2} &= \tfrac14(1+\lambda_z \pm (\lambda_x+\lambda_y)), \nonumber \\
p_{3,4} &= \tfrac14(1-\lambda_z \pm (\lambda_x-\lambda_y)).
\end{align}

In concluding this subsection, it is worth to mention that a sufficient condition for Bell-diagonal states to saturate the lower bound in the MA-EUR relation has been obtained in Ref.~\cite{xu12-2}. Since $\varrho_{AB}$ retains its Bell-diagonal form under the action of Pauli channels, it could be expressed in our treatment as $\lambda_y=\lambda_x \lambda_z$. In other words, provided that this condition is satisfied, then $U_{\text{su}}=B_{\text{su}}$. This is indeed straightforward to see by directly substituting $\lambda_y=\lambda_x \lambda_z$ in the four eigenvalues $p_k$ in Eq.~\eqref{eigvalues}, which represent a joint probability distribution, and observing that they can be factorized into two independent distributions. When this condition is written in terms of the Pauli channel parameters in Eq.~\eqref{PauliChan} utilizing Eq.~\eqref{lambdas}, it becomes $q_0q_y=q_xq_z$. Equivalently, in terms of the bias parameters $\alpha_x,\alpha_y,\alpha_z$ and the overall error probability $p$, the saturation condition reads (for $p\neq 0$)
\begin{equation} \label{boundcond}
\alpha_y(1-p)=p\alpha_x\alpha_z,
\end{equation}
which is satisfied for all $p\in(0,1]$ provided that 
\begin{equation}
\alpha_y=0, \quad \alpha_x \alpha_z =0, \quad \alpha_x+\alpha_z=1,
\end{equation} 
which solely holds for the bit-flip ($\alpha_x=1, \alpha_y=\alpha_z=0$) and the phase-flip ($\alpha_x=\alpha_y=0,\alpha_z=1$) channels.

\subsection{Self-switched Pauli Channels}

Let us consider the self-switching of two identical Pauli channels $\Lambda_1=\Lambda_2$, where the Kraus operators for each of them are given as $M_i=\sqrt{q_i}\sigma_i$, where $i,j \in \{0,x,y,z\}$. Then, the Kraus operators of the self-switch channel $\mathcal{S}$ given in Eq.~\eqref{kraussw} can be expressed as
\begin{equation}
S_{ij} = |0\rangle\langle 0|_A \otimes \sqrt{q_i q_j}\sigma_i\sigma_j + |1\rangle\langle 1|_A \otimes \sqrt{q_i q_j}\sigma_j\sigma_i.
\end{equation}
In accordance, under the action of the self-switch channel $\mathcal{S}$ in Eq.~\eqref{sschan}, the density operator $\varrho_{AB}$ is mapped to
\begin{align}
\varrho_{AB}^{\text{sw}} ={}& \tfrac14 [\mathcal{S}(\mathbb{I}\otimes \mathbb{I}) + \mathcal{S}(\sigma_x\otimes \sigma_x) \nonumber \\
& - \mathcal{S}(\sigma_y\otimes \sigma_y) + \mathcal{S}(\sigma_z\otimes \sigma_z)],
\end{align}
where the superscript sw denotes switch. We note that $\mathcal{S}$ is constructed from the Pauli channels $\Lambda$, and one of the characteristic properties of them is that they transform Pauli operators onto themselves up to a scalar multiple. To put it differently, the Pauli operators are the eigenvectors of the Pauli channels. As a consequence, the action of the self-switch channel $\mathcal{S}$ on the Bell state $\varrho_{AB}$ given in Eq.~\eqref{Bellstate} results in a density operator preserving the Bell-diagonal form. That is, the channel $\mathcal{S}$ maps each of the operators in Eq.~\eqref{Bellstate} back to themselves, multiplied by some new coefficient $\kappa_\mu$ with $\mu \in \{x,y,z\}$, leading to
\begin{align} \label{evolvedrhosw}
\varrho_{AB}^{\text{sw}} ={}& \tfrac14 [(\mathbb{I}\otimes \mathbb{I}) + \kappa_x(\sigma_x\otimes \sigma_x) \nonumber \\
& - \kappa_y(\sigma_y\otimes \sigma_y) + \kappa_z(\sigma_z\otimes \sigma_z)].
\end{align}
Using the algebra of Pauli operators, the parameters $\kappa_\mu$ are calculated in Appendix~\ref{app-a} as 
\begin{align} 
\kappa_x ={}& \big(1 - 2p(\alpha_y+\alpha_z)\big)^2 \nonumber \\
  &+ 4p^2\big(\alpha_x\alpha_y + \alpha_x\alpha_z - \alpha_y\alpha_z\big),  \nonumber \\[1ex] 
\kappa_y ={}& \big(1 - 2p(\alpha_x+\alpha_z)\big)^2 \nonumber \\
  &+ 4p^2\big(\alpha_x\alpha_y - \alpha_x\alpha_z + \alpha_y\alpha_z\big),  \nonumber  \\[1ex] 
\kappa_z ={}& \big(1 - 2p(\alpha_x+\alpha_y)\big)^2 = \lambda_z^2.
\end{align}

Let us turn our attention to the calculation of the conditional entropies in MA-EUR for the self-switch channel. We recall that in our treatment while the measured system $A$ in the MA-EUR also plays the role of the control for the quantum switch, the memory $B$ is affected by a noise described through self-switching of Pauli channels. As the density operator $\varrho_{AB}^{\text{sw}}$ in Eq.~\eqref{evolvedrhosw} has exactly the same form as the density operator $\varrho_{AB}^{\text{su}}$ in Eq.~\eqref{evolvedrhosu}, calculation of conditional entropies is straightforward. Particularly, the density operators after the non-selective measurements in $\sigma_x$ and $\sigma_z$ bases are given by~\cite{mynote}
\begin{align}
\varrho_{XB}^{\text{sw}} &= \tfrac14 \left[\mathbb{I}\otimes  \mathbb{I} + \kappa_x(\sigma_x\otimes \sigma_x)\right], \nonumber \\
\varrho_{ZB}^{\text{sw}} &=  \tfrac14 \left[\mathbb{I}\otimes  \mathbb{I} + \lambda^2_z(\sigma_z\otimes \sigma_z)\right].
\end{align}
Therefore, we obtain the conditional entropies as 
\begin{align}
S(X|B)_{\text{sw}}&=S(\varrho_{XB}^{\text{sw}})-S(\varrho_B^{\text{sw}}) = H_\text{bin}\Big( \frac{1+\kappa_x}{2}\Big), \\
S(Z|B)_{\text{sw}}&=S(\varrho_{ZB}^{\text{sw}})-S(\varrho_B^{\text{sw}}) = H_\text{bin}\Big( \frac{1+\lambda^2_z}{2}\Big),
\end{align}
which results in a total entropic uncertainty,
\begin{align} \label{Usw}
U_{\text{sw}} = H_\text{bin}\Big( \frac{1+\kappa_x}{2}\Big) +  H_\text{bin}\Big( \frac{1+\lambda^2_z}{2}\Big).
\end{align}
On the other hand, the uncertainty bound in this case is
\begin{equation}\label{BSw}
B_{\text{sw}}=S(\varrho_{AB}^{\text{sw}})=-\sum\nolimits_k r_k \log_2 r_k,
\end{equation}
where $r_k$ are the four eigenvalues of the bipartite density operator $\varrho_{AB}^{\text{sw}}$, which can be found to be
\begin{align} \label{eigvalueskap}
r_{1,2} &= \tfrac14(1+\lambda^2_z \pm (\kappa_x+\kappa_y)), \nonumber \\
r_{3,4} &= \tfrac14(1-\lambda^2_z\pm (\kappa_x-\kappa_y)),
\end{align}
which completes the calculation of all the terms appearing on both sides of the MA-EUR.

Now that we have found the total entropic uncertainty in both single-use and self-switch cases for Pauli channels, respectively denoted by $U_{\text{su}}$ and $U_{\text{sw}}$, we are ready to investigate the conditions under which self-switch of Pauli channels reduce the total uncertainty that Bob has about the measurement outcomes of Alice, as compared to their direct application on the memory qubit $B$ of Bob. To determine the parameter region where the switch provides an advantage in reducing uncertainty, we compare the total uncertainties derived in Eq.~\eqref{Usu} and in Eq.~\eqref{Usw}. Clearly, self-switching of Pauli channels reduce the total entropic uncertainty as compared to the single-use case when $U_{\text{sw}}<U_{\text{su}}$, which
explicitly reads
\begin{equation} \label{adv}
h(\kappa_x) + h(\lambda^2_z) < h(\lambda_x) + h(\lambda_z), 
\end{equation}
where $h(t)\equiv H_\text{bin}[(1+t)/2]$. Note that $\lambda_\mu\in[-1,1]$ and $h(t)$ is even and strictly decreasing in $|t|$ for $t\in[-1,1]$. Thus, we have $\lambda^2_z \leq |\lambda_z|$, which implies that the entropic uncertainty for $\sigma_z$ measurement, $S(Z|B)_\text{sw}$, for switched channel is greater than or equal to that of single-use case, that is, $h(\lambda_z) \leq h(\lambda^2_z)$. Rearranging Eq.~\eqref{adv}, we write
\begin{equation} \label{adv2}
h(\lambda^2_z) - h(\lambda_z) < h(\lambda_x) - h(\kappa_x).
\end{equation}
As a consequence, since the left-hand side is always non-negative, a reduction in total uncertainty is only possible if the self-switch channel $\mathcal{S}$ sufficiently diminishes the uncertainty for $\sigma_x$ measurement, $S(X|B)_\text{sw}$, which requires that $h(\kappa_x)<h(\lambda_x)$. This in turn implies that a reduction in x-uncertainty demands that $|\lambda_x| < |\kappa_x|$. As we are not able to obtain an algebraic expression guaranteeing that Eq.~\eqref{adv2}, which is a necessary and sufficient condition for the quantum switch based reduction of the total uncertainty, is satisfied, to obtain an intuitive result, we rather focus on the necessary condition $|\lambda_x| < |\kappa_x|$, ensuring a decrease in x-uncertainty.
\begin{figure}[t]
\centering
\includegraphics[width=0.48\textwidth]{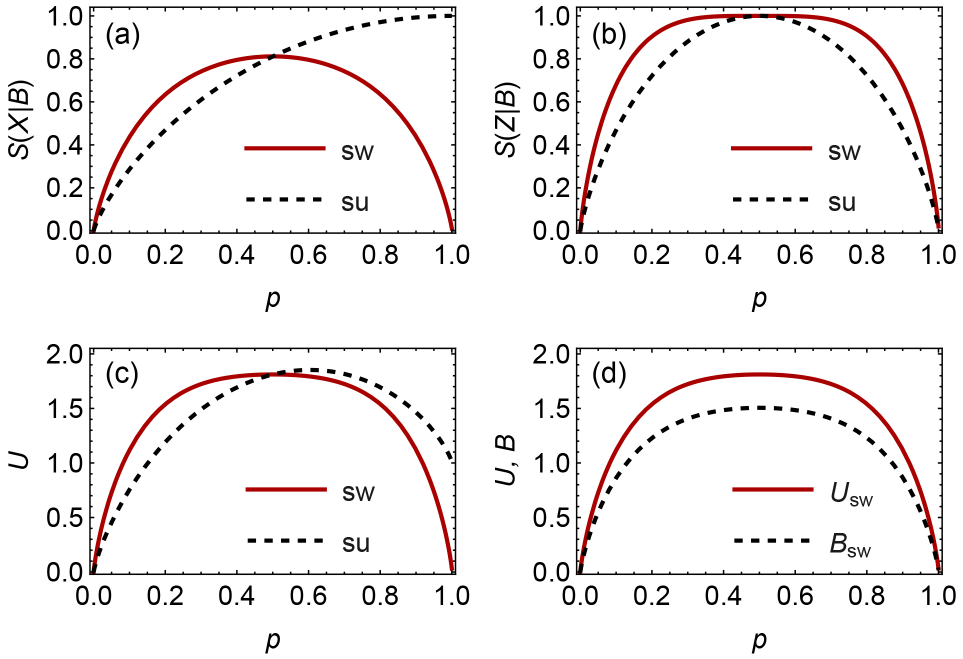}
\caption{Assuming that $\alpha_x=\alpha_y=0.5$ and $\alpha_z=0$, Bob's uncertainty about x-measurement (a) and z-measurement (b) for switched (sw) and single-use (su) Pauli channels in terms of the overall error probability $p$. (c) Bob's total uncertainty for switched $U_{\text{sw}}$ and single-use $U_{\text{su}}$ channels. (d) Total uncertainty $U_{\text{sw}}$ and its lower bound $B_{\text{sw}}$ for switched channels.} 
\label{fig1}
\end{figure}
If we suppose that $\lambda_x,\kappa_x \geq 0$, then a rather straightforward calculation shows that
\begin{equation} \label{adv3}
p>\frac{\alpha_y+\alpha_z}{2(\alpha_y+\alpha_z-\alpha_y\alpha_z)}.
\end{equation}
It is not difficult to realize that the minimum value of the right-hand side is 0.5, which is attained when $\alpha_y\alpha_z=0$. That is, a necessary condition for switch based advantage is given by Eq.~\eqref{adv3}, and it always necessitates $p>1/2$. At this point we should stress that, in obtaining Eq.~\eqref{adv3}, we have assumed that $\lambda_x$ and $\kappa_x$ are both non-negative. It is also possible for these parameters to have different signs, which might lead to a different necessary condition than the expression stated in Eq.~\eqref{adv3}. Nonetheless, we omit such cases for now to concentrate on $\lambda_x,\kappa_x>0$ case to give concrete examples, where the total uncertainty is reduced via the quantum switch. Following Eq.~\eqref{adv2}, this net reduction solely occurs if the amount of the decrease in x-uncertainty overcomes the increase in z-uncertainty.

\begin{figure}[t]
\centering
\includegraphics[width=0.48\textwidth]{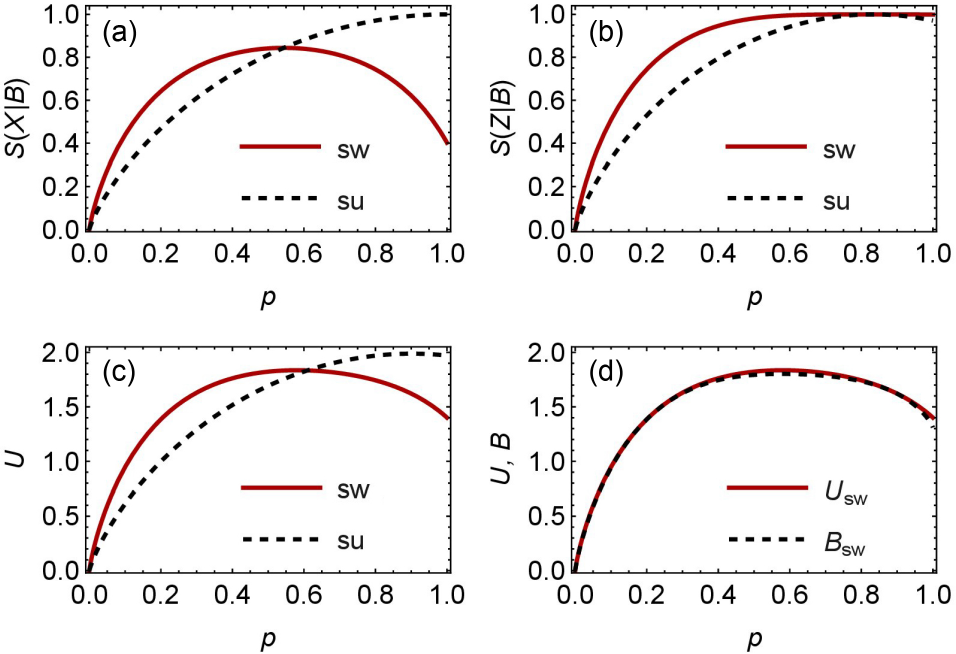}
\caption{Supposing $\alpha_x=0.5$, $\alpha_y=0.1$ and $\alpha_z=0.4$, Bob's uncertainty about x-measurement (a) and z-measurement (b) for switched (sw) and single-use (su) Pauli channels in terms of the overall error probability $p$. (c) Bob's total uncertainty for switched $U_{\text{sw}}$ and single-use $U_{\text{su}}$ channels. (d) Total uncertainty $U_{\text{sw}}$ and its lower bound $B_{\text{sw}}$ for switched channels.} 
\label{fig2}
\end{figure}

We first consider the Pauli channel having the bias parameters $\alpha_x=\alpha_y=0.5$ and $\alpha_z=0$, which represents an equal combination of bit-flip and bit-phase-flip errors. In this case, we notice that $\lambda_x,\kappa_x\geq0$ for all $p \in [0,1]$. Setting the bias parameter $\alpha_z=0$ ensures that the right-hand side of Eq.~\eqref{adv3} acquires its minimum possible value. Then, when the threshold is crossed, that is once $p>0.5$, the entropic uncertainty for x-measurement $S(X|B)$ is always smaller for the self-switched Pauli channels as compared to the single-use case. Furthermore, for this particular Pauli channel, the condition in Eq.~\eqref{adv3} turns out to be not only necessary but also sufficient for reducing the total uncertainty, due to strict monotonicity of the binary entropy. We display the uncertainty that Bob has regarding x and z measurements on Alice's qubit, that is, $S(X|B)$ and $S(Z|B)$, in Fig.~\ref{fig1}(a) and Fig.~\ref{fig1}(b) in cases of self-switched and single-use Pauli channels as a function of the overall error probability $p$. Besides, Fig.~\ref{fig1}(c) shows Bob's total uncertainty $U_{\text{sw}}$ and $U_{\text{su}}$ about these two measurements for self-switched and single-use Pauli channels, respectively. It can be seen from Fig.~\ref{fig1}(a) and Fig.~\ref{fig1}(c) that once $p>0.5$ both the x-measurement uncertainty and the total uncertainty for the self-switched Pauli channel simultaneously become lower than that of the single-use Pauli channel, which confirms that the condition in Eq.~\eqref{adv3} is actually necessary and sufficient in this case for reducing total uncertainty through switch. In addition, Fig.~\ref{fig1}(b) demonstrates that the uncertainty in z-measurement is never lower for the switched channel than for the single-use channel. Lastly, in Fig.~\ref{fig1}(d), we show the total uncertainty for the switched channel $U_{\text{sw}}$ along with its lower bound $B_{\text{sw}}$. We also emphasize that self-switching this channel inverts the noise after $p>0.5$, eventually restoring the initial Bell state $\varrho_{AB}$, and thus making the total uncertainty vanish again at $p=1$. 

Let us next consider another channel with the bias parameters $\alpha_x=0.5$, $\alpha_y=0.1$, and $\alpha_z=0.4$ that represents a generic Pauli channel with no particular symmetry. Note that we again have $\lambda_x,\kappa_x\geq0$ for all $p \in [0,1]$. Fig.~\ref{fig2} shows the results of a similar analysis as in Fig.~\ref{fig1} for this generic channel. We can observe from Fig.~\ref{fig2}(c) that self-switching Pauli channels provides a total uncertainty advantage for these bias parameters as well. In this case, the x-uncertainty becomes smaller for the switched Pauli channel than the single-use channel when $p\gtrsim0.54$. However, the advantage in total uncertainty only appears once $p\gtrsim0.6$ since $U_{\text{sw}}$ becomes lower than $U_{\text{su}}$ after that point, demonstrating that Eq.~\eqref{adv3} is only necessary but not sufficient in general for switch based advantage.
\begin{figure}[t]
\centering
\includegraphics[width=0.48\textwidth]{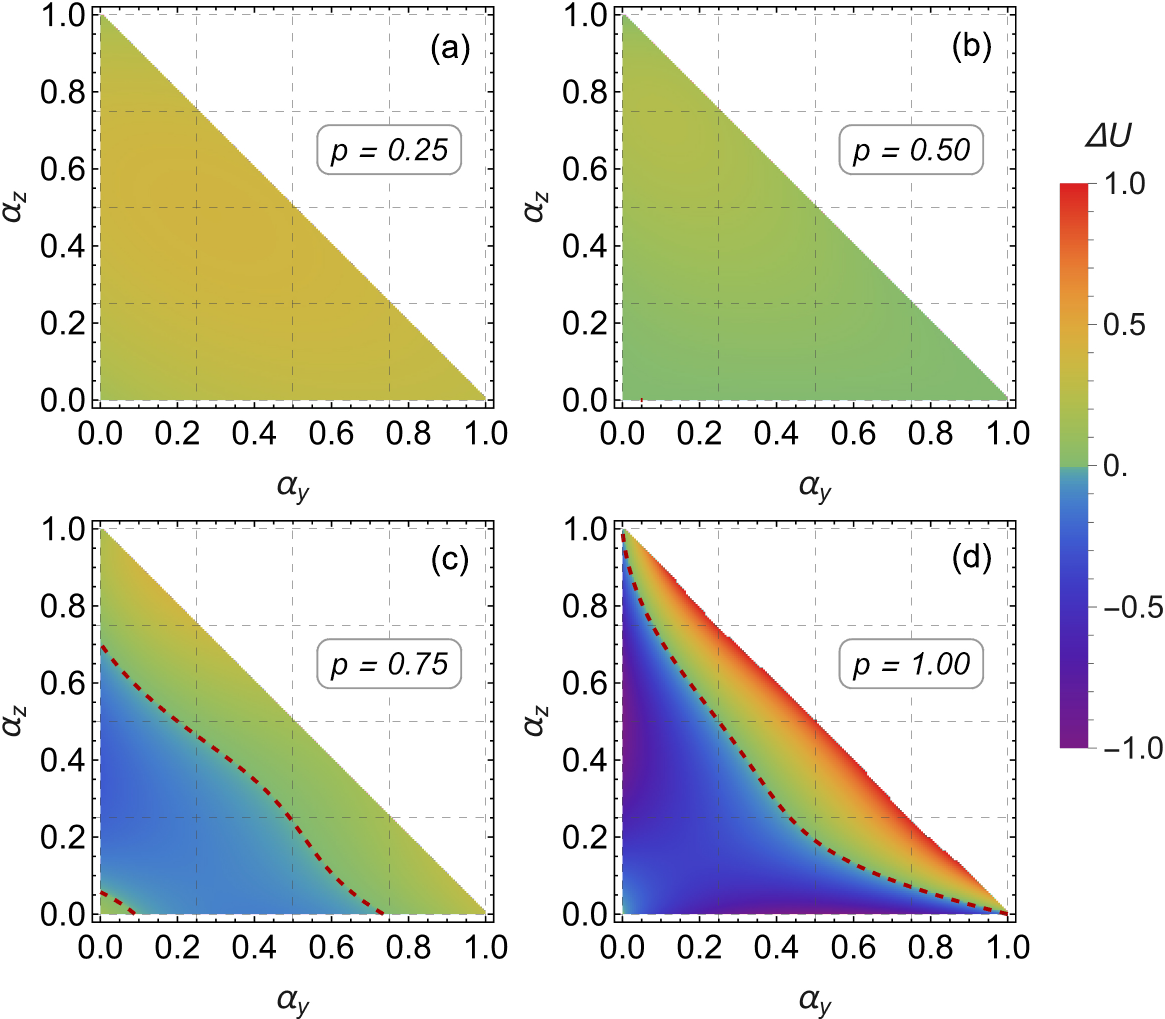}
\caption{Density plots of $\Delta U = U_\text{sw}-U_\text{su}$, i.e., the difference between the total entropic uncertainties for the self-switched and the single-use Pauli channels. White regions in the panels correspond to the non-physical regime, where $\alpha_x+\alpha_y+\alpha_z=1$ with $\alpha_x,\alpha_y,\alpha_z\geq0$ is not satisfied. There exists no switch based advantage for the total uncertainty when $p=0.25$ (a) and $p=0.50$ (b) as $\Delta U$ is always positive for Pauli channels. The panels in (c) and (d) display the regions where $\Delta U<0$, and thus where the total entropic uncertainty of Bob about Alice's measurements is decreased due to the self-switching of Pauli channels, respectively, for $p=0.75$ and $p=1.00$. Thick dashed lines correspond to the contours on which $U_\text{sw}= U_\text{su}$.}
\label{fig3}
\end{figure}

Before we conclude this subsection, we want to present a complete numerical analysis of the necessary and sufficient condition for self-switch based advantage, expressed in Eq.~\eqref{adv2}, considering the full family of Pauli channels. Fig.~\ref{fig3} summarizes the outcomes of our numerical analysis displaying the parameter region in which switch based advantage manifests. Here we observe that there exits no reduction of total uncertainty due to switch when $p<0.5$. Recall that we supposed $\lambda_x,\kappa_x>0$ to obtain the necessary condition in Eq.~\eqref{adv3} and  then showed that $\Delta U <0$ requires $p>0.5$. Now, with Fig.~\ref{fig3}, we can observe that this is still true even when $\lambda_x$ and $\kappa_x$ have different signs.

\subsection{Time-flipped Pauli Channels}

We now consider the quantum time-flip for Pauli channels. Similarly to the previous subsection, initial state of the bipartite system $AB$ is assumed to be the Bell state represented by the density operator $\varrho_{AB}=|\psi\rangle \langle \psi|$. The Kraus operators describing time-flip channel $\mathcal{F}$ are
\begin{equation}
F_{i} = |0\rangle\langle 0|_A \otimes \sqrt{q_i}\sigma_i\ + |1\rangle\langle 1|_A \otimes \sqrt{q_i}\sigma_i^T,
\end{equation}
where $i \in \{0,x,y,z\}$. Under the action of the time-flip channel $\mathcal{F}$ in Eq.~\eqref{tfchann}, the density operator $\varrho_{AB}$ becomes
\begin{align}
\varrho_{AB}^{\text{tf}} ={}& \tfrac14 [\mathcal{F}(\mathbb{I}\otimes \mathbb{I}) + \mathcal{F}(\sigma_x\otimes \sigma_x) \nonumber \\
& - \mathcal{F}(\sigma_y\otimes \sigma_y) + \mathcal{F}(\sigma_z\otimes \sigma_z)],
\end{align}
where the superscript tf denotes time-flip. The time-flip channel $\mathcal{F}$  preserves the Bell-diagonal structure as well. In particular, $\mathcal{F}$ maps the Bell state $\varrho_{AB}$ in Eq.~\eqref{Bellstate} into
\begin{align} \label{evolvedrhotf}
\varrho_{AB}^{\text{tf}} ={}& \tfrac14 [(\mathbb{I}\otimes \mathbb{I}) + \tau_x(\sigma_x\otimes \sigma_x) \nonumber \\
& - \tau_y(\sigma_y\otimes \sigma_y) + \tau_z(\sigma_z\otimes \sigma_z)],
\end{align}
where the parameters $\tau_\mu$ are derived in Appendix~\ref{app-b} as
\begin{gather} \label{taus} 
\tau_x ={} 1 - 2\alpha_zp,  \qquad \tau_y ={} 1 - 2p, \nonumber \\[1ex]  
\tau_z ={} 1 - 2(1-\alpha_z)p = \lambda_z.
\end{gather}
The post-measurement states after non-selective measurements are performed in $\sigma_x$ and $\sigma_z$ bases on $\varrho_{AB}^{\text{tf}}$ read
\begin{align}
\varrho_{XB}^{\text{tf}} &= \tfrac14 \left[\mathbb{I}\otimes  \mathbb{I} + \tau_x(\sigma_x\otimes \sigma_x)\right], \nonumber \\
\varrho_{ZB}^{\text{tf}} &=  \tfrac14 \left[\mathbb{I}\otimes  \mathbb{I} + \lambda_z(\sigma_z\otimes \sigma_z)\right].
\end{align}
Hence, the conditional entropies can be written as
\begin{align}
S(X|B)_{\text{tf}}&=S(\varrho_{XB}^{\text{tf}})-S(\varrho_B^{\text{tf}}) = h( \tau_x ), \\
S(Z|B)_{\text{tf}}&=S(\varrho_{ZB}^{\text{tf}})-S(\varrho_B^{\text{tf}}) = h( \lambda_z ),
\end{align}
which results in a total entropic uncertainty,
\begin{align} \label{Utf}
U_{\text{tf}} = h( \tau_x ) + h( \lambda_z)
\end{align}
Besides, the uncertainty bound can be calculated as
\begin{equation}\label{Btf}
B_{\text{tf}}=S(\varrho_{AB}^{\text{tf}})=-\sum\nolimits_k s_k \log_2 s_k,
\end{equation}
where $s_k$ are the four eigenvalues of the bipartite density operator $\varrho_{AB}^{\text{tf}}$, that can be expressed as
\begin{align} \label{eigvaluestau}
s_{1,2} &= \tfrac14(1+\lambda_z \pm (\tau_x+\tau_y)), \nonumber \\
s_{3,4} &= \tfrac14(1-\lambda_z\pm (\tau_x-\tau_y)),
\end{align}

\begin{figure}[t]
\centering
\includegraphics[width=0.48\textwidth]{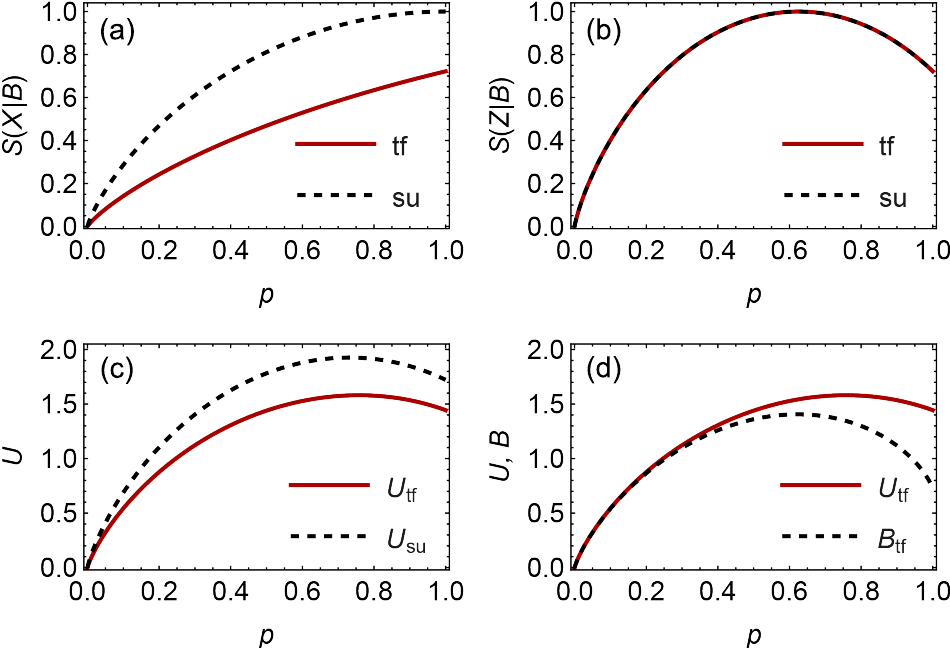}
\caption{For $\alpha_x=0.5$, $\alpha_y=0.3$ and $\alpha_z=0.2$, Bob's uncertainty about x-measurement (a) and z-measurement (b) for time-flipped (tf) and single-use (su) Pauli channels in terms of the overall error probability $p$. (c) Bob's total uncertainty for time-flipped $U_{\text{tf}}$ and single-use $U_{\text{su}}$ channels. (d) Total uncertainty $U_{\text{tf}}$ and its lower bound $B_{\text{tf}}$ for time-flipped channels.} 
\label{fig4}
\end{figure}

At this point, our aim is to explore the conditions under which time-flip based advantage in reducing the total uncertainty manifests for Pauli channels. It is clear that such an advantage is ensured when $U_{\text{tf}}<U_{\text{su}}$ is satisfied. More explicitly, the condition given by
\begin{equation} \label{advtf}
h(\tau_x) + h(\lambda_z) < h(\lambda_x) + h(\lambda_z), 
\end{equation}
is necessary and sufficient for time-flip based advantage and it simply reduces to $h(\tau_x) < h(\lambda_x)$, which is in turn equivalent to $|\tau_x|>|\lambda_x|$. Considering the four different sign cases for $\tau_x$ and $\lambda_x$, it can be shown that (for $p>0$)
\begin{equation} \label{advtf2}
\quad 0 < \alpha_y p < 1 - 2\alpha_z p,
\end{equation}
provides a necessary and sufficient condition for a net decrease in total entropic uncertainty for time-flipped Pauli channels in comparison with the single-use case. We note that, as opposed to the self-switch case, here we are able to obtain an algebraic necessary and sufficient condition thanks to the fact that z-measurement uncertainty is the same for both single use and time-flipped Pauli channels.

In Fig.~\ref{fig4}, we exemplify the situation for the Pauli channel defined by $\alpha_x=0.5$, $\alpha_y=0.3$ and $\alpha_z=0.2$. Fig.~\ref{fig4}(a) and Fig.~\ref{fig4}(b) illustrate that while Bob's x-measurement uncertainty is lower for the time-flipped channel than for the single-use case for all $p\in(0,1]$, his z-measurement uncertainties regarding Alice's outcomes are identical for both single-use and flipped channels. It is easy to confirm that the bias parameters defining this particular channel satisfy the condition given in Eq.~\eqref{advtf2} for time-flip based advantage in reducing the total uncertainty. This implies that the total uncertainties for the time-flip and single-use cases, namely $U_\text{tf}$ and $U_\text{su}$, satisfies $U_\text{tf}<U_\text{su}$ when $0<p\leq1$, which is clearly demonstrated in Fig.~\ref{fig4}(c). Besides, in Fig.~\ref{fig4}(d), we also show the total uncertainty $U_\text{tf}$ along with its lower bound $B_\text{tf}$ for the time-flip channel.

\begin{figure}[t]
\centering
\includegraphics[width=0.48\textwidth]{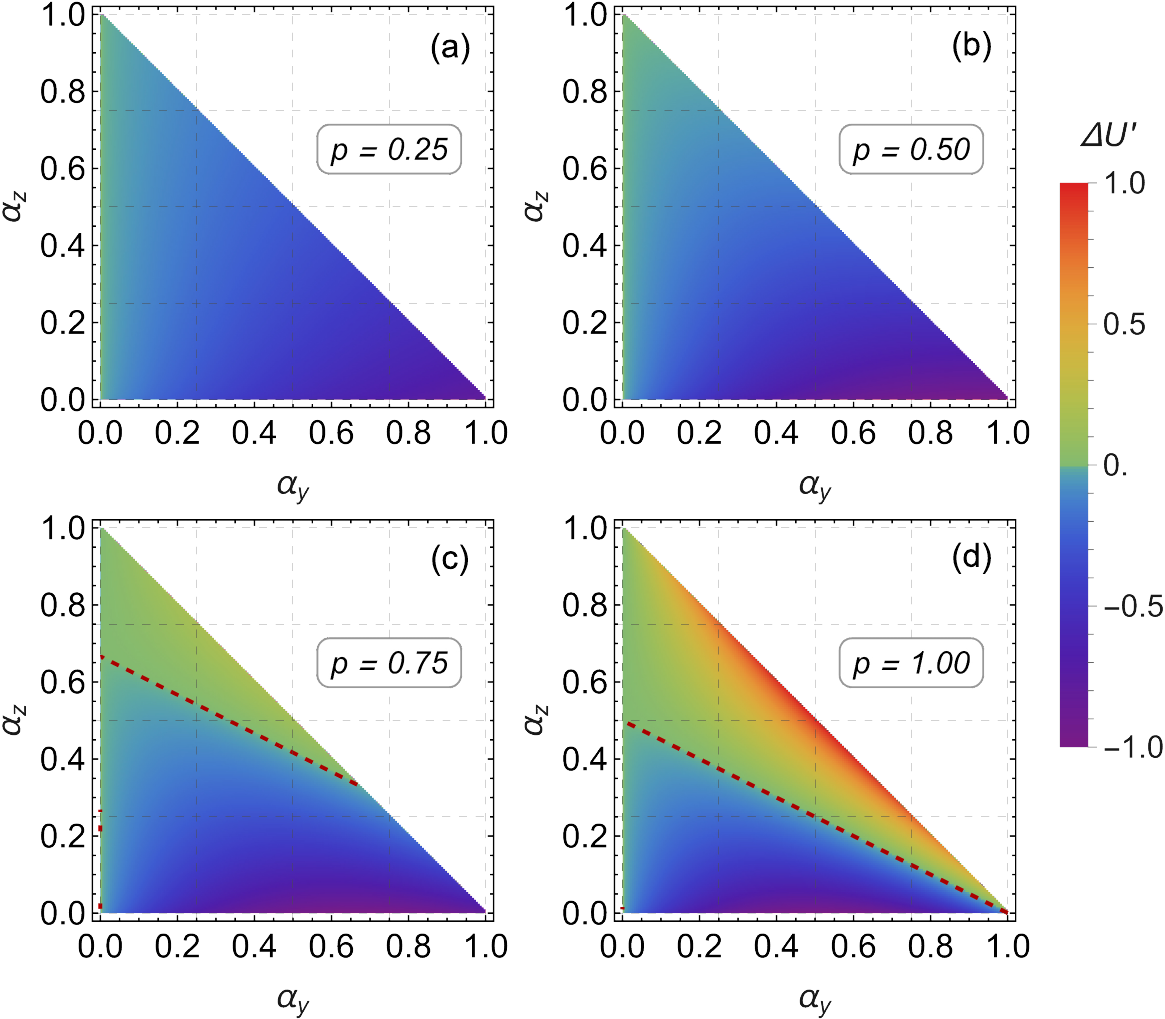}
\caption{Density plots of $\Delta U' = U_\text{tf}-U_\text{su}$, i.e., the difference between the total entropic uncertainties for the time-flipped and the single-use Pauli channels. White regions in the panels correspond to the non-physical regime, where $\alpha_x+\alpha_y+\alpha_z=1$ with $\alpha_x,\alpha_y,\alpha_z\geq0$ is not satisfied. There always exists time-flip based advantage (except for $\alpha_y=0$) for total uncertainty when $p=0.25$ (a) and $p=0.50$ (b) as $\Delta U'$ is always negative. The panels in (c) and (d) display the regions where $\Delta U'<0$, and thus where the total uncertainty of Bob about Alice's measurements is decreased due to the time-flipped Pauli channels, respectively, for the cases $p=0.75$ and $p=1.00$. Thick dashed lines correspond to the contours on which $U_\text{tf}= U_\text{su}$.}
\label{fig5}
\end{figure}

Lastly, we consider the complete family of Pauli channels, characterized by $\alpha_x$, $\alpha_y$ and $\alpha_z$, to display the region where the total entropic uncertainty is diminished through the implementation of time-flip superchannel in comparison with the single-use case. This comparison benchmarks the time-flip against the ordinary single use of the same noise channel, mapping the specific regimes where the implementation of indefinite input-output direction provides a genuine advantage in reducing the total entropic uncertainty. More explicitly, the panels in Fig.~\ref{fig5} display the regions where $\Delta U'=U_\text{tf}-U_\text{su}<0$ for four different error probabilities. In each panel, these regions are analytically determined by the condition in Eq.~\eqref{advtf2}. Hence, there is a wide class of Pauli channels, for which the time-flip provides an advantage in reducing the total uncertainty. In particular, the greatest advantage is attained, provided that $\alpha_z=0$ and $\alpha_y=\min(1,1/2p)$, in which case $\Delta U'=-1$ when $\alpha_y=1/2p\leq1$.

\section{Conclusion and Discussion} \label{sec5}

In summary, we examined how higher-order controlled processes, known as the quantum switch and the quantum time-flip, impact the memory-assisted entropic uncertainty for Pauli noise acting on the quantum memory. In the operational setting we considered, the same system that implements the coherent control of quantum channels is the very system probed by the MA-EUR, while the noise acts on the memory. Our analysis shows that both the quantum switch and the quantum time-flip can reduce the total entropic uncertainty relative to the single-use action of the Pauli channels on the memory.

In case of self-switched channels, we demonstrated that reductions in total entropic uncertainty appear beyond a certain noise threshold for a broad class of Pauli channels. We obtained a necessary analytic condition for such an advantage to manifest, which becomes also sufficient for some specific channels. In addition, performing a full numerical scan over the channel parameter space, we confirmed that indeed no reduction occurs at low noise for any Pauli channel, and that such an advantage in reducing the uncertainty grows once the threshold is crossed. On the other hand, for the time-flipped Pauli channels, we derived a simple necessary and sufficient condition for reduced uncertainty relative to direct single-use action. In this case, advantage can emerge at any noise strength, depending on the bias parameters of the channel.

The results presented here can be understood from a common structural property shared by both higher-order quantum processes. In cases of both the quantum switch and the time-flip, the joint output state of the control qubit $A$ and the memory $B$ has a block form in the control basis, where the off-diagonal blocks encode the coherent combination of the two underlying alternatives. When the qubit $A$ is measured in a basis that preserves coherence, e.g., $\sigma_x$, these off-diagonal contributions affect the conditional states of qubit $B$ and thus the conditional entropies $S(Q|B)$ and $S(R|B)$. However, measuring $A$ in an incoherent basis, e.g., $\sigma_z$, suppresses these contributions. The conceptual difference lies in what alternatives are combined, that is, whereas the  quantum switch combines the two causal orders of the channels, the quantum time-flip combines forward versus input-output inverted realizations of the same dynamics. This is basically why both can yield uncertainty reduction while producing distinct advantage landscapes.

Before we conclude, a few last remarks are in order. To start with, it is important to discuss the dependence of our results on the choice of the measurements. In our treatment, we focused on the observables $\sigma_x$ and $\sigma_z$ for two main reasons. First, they represent the two limiting cases of the interference mechanism. On the one hand, $\sigma_z$ is diagonal in the control basis and thus it is effectively blind to the off-diagonal interference terms contributing to the dynamics. On the other hand, $\sigma_x$ is purely off-diagonal in the control basis, which lets us fully access the interference terms. Second, this choice renders the dynamics analytically tractable, allowing us to derive exact closed-form conditions for the advantage regions, which would be difficult to obtain for arbitrary observable pairs. While using other incompatible measurement pairs would generally change the quantitative boundaries and thresholds of the advantage regions, the physical mechanism responsible for the reduction of the total entropic uncertainty remains unchanged. Therefore, our findings can be regarded as a representative case study using the canonical observable pair that establishes the physical mechanism of the advantage rather than demonstrating it for a specific choice of measurement pair.

Furthermore, we comment on the dependence of the advantage on the purity of the initial bipartite state. Although we consider a maximally entangled Bell state in our calculations, the same analysis extends straightforwardly to the broad and physically relevant class of Bell-diagonal mixed states, which can be written as
\begin{equation} 
\varrho_{AB}
=\frac{1}{4}\Bigl(\mathbb{I}\otimes \mathbb{I}+\sum_{j=x,y,z} c_j\,\sigma_j\otimes\sigma_j\Bigr),
\end{equation}
where the coefficients $(c_x,c_y,c_z)$ satisfy the Bell-diagonal positivity constraints~\cite{Horodecki13}. For this family of density operators, the local Pauli channel and the corresponding quantum switch and time-flip constructions built from it preserve the Bell-diagonal structure, and their effect is simply to rescale the coefficients. In the single-use case, the output state coefficients will be scaled as $c_j\lambda_j$, while for the quantum switch and the quantum time-flip they take the form $c_j\kappa_j$ and $c_j\tau_j$, respectively. Consequently, the conditional entropies entering the total entropic uncertainty, $U=S(Q|B)+S(R|B)$, retain the same closed-form expressions as in the Bell state case, with the effective coefficients simply replaced by $c_j\lambda_j$, $c_j\kappa_j$, and $c_j\tau_j$ in the single-use, switch, and time-flip cases, respectively. Thus, the results obtained for the maximally entangled state extend to the broader class of Bell-diagonal states, confirming that the reported advantages reflect a generic feature of the dynamics rather than a consequence of the chosen pure initial state, even though the quantitative advantage landscape is naturally determined by the degree of initial mixedness. An interesting research direction is to investigate whether similar conclusions hold for more general noise models beyond Pauli channels (including unital but non--Pauli-diagonal channels or non-unital dissipative channels), where Bell-diagonal structure need not remain preserved and new behavior may emerge.

\begin{acknowledgments}
G.K. is supported by The Scientific and Technological Research Council of T\"{u}rkiye (TUBITAK) through the 100th Anniversary Incentive Award. The author would like to thank B. Ç. for his comments on the manuscript.
\end{acknowledgments}

\appendix
\section{Derivation of the state parameters \texorpdfstring{$\kappa_x, \kappa_y$ and $\kappa_z$}{kappa-x, kappa-y and kappa-z} under self-switched Pauli channels} \label{app-a}

First of all, the reason that the first term in Eq.~\eqref{Bellstate} is invariant under $\mathcal{S}$ is that the Pauli channels are unital. Next, $\kappa_z$ can also be determined in a straightforward way, recognizing that the action of the self-switch channel $\mathcal{S}$ on $\sigma_z\otimes \sigma_z$ is determined by the diagonal superoperators $\mathcal{S}_{00}$ and $\mathcal{S}_{11}$, which follows from Eq.~\eqref{sup-mat}. In particular, 
\begin{align}
\mathcal{S}(\sigma_z \otimes \sigma_z) ={}& |0\rangle\langle0| \otimes \mathcal{S}_{00}(\sigma_z) + |1\rangle\langle1| \otimes \mathcal{S}_{11}(-\sigma_z), \nonumber \\
={}& |0\rangle\langle0| \otimes \mathcal{S}_{00}(\sigma_z) - |1\rangle\langle1| \otimes \mathcal{S}_{11}(\sigma_z)
\end{align}
Noting that two identical Pauli channels are switched in our treatment, we have $\mathcal{S}_{00}=\mathcal{S}_{11}$ and hence the action of this mapping is equivalent to the twice application of the Pauli channel on the second qubit in series, i.e.,
\begin{align}
\mathcal{S}(\sigma_z \otimes \sigma_z) &= \sigma_z\otimes\mathcal{S}_{00}(\sigma_z) = \sigma_z\otimes \Lambda^2 (\sigma_z) \nonumber \\
&= \sigma_z\otimes \Lambda (\lambda_z\sigma_z) = \lambda_z^2(\sigma_z \otimes \sigma_z),
\end{align}
which clearly leads to the conclusion that $\kappa_z=\lambda_z^2$. Moving forward to determine $\kappa_x$, we note that 
\begin{equation}
\mathcal{S}(\sigma_x\otimes \sigma_x)= |0\rangle\langle1| \otimes \mathcal{S}_{01}(\sigma_x) + |1\rangle\langle0| \otimes \mathcal{S}_{10}(\sigma_x),
\end{equation}
which is due to the structure of the switched channel in Eq.~\eqref{sup-mat} and the block off-diagonal form of $\sigma_x\otimes \sigma_x$. As identical Pauli channels are fed into the quantum switch, we also have $\mathcal{S}_{01}=\mathcal{S}_{10}$ and thus
\begin{equation} 
\mathcal{S}(\sigma_x\otimes \sigma_x)= \sigma_x \otimes \mathcal{S}_{01}(\sigma_x).
\end{equation}
Unlike the case of $\kappa_z$, here we need to explicitly calculate the action of the superoperator $\mathcal{S}_{01}$ on $\sigma_x$ as follows:
\begin{equation}\label{super01}
\mathcal{S}_{01}(\sigma_x) = \sum_{i,j} q_iq_j(\sigma_i\sigma_j) \sigma_x (\sigma_i \sigma_j),
\end{equation}
where $i,j \in \{0,x,y,z\}$. We sum over the sixteen terms to obtain $\kappa_x$, noting that each term in this sum maps $\sigma_x$ to itself up to a multiplicative constant written in terms of the channel parameters $q_i$'s. Let $\Omega=\sigma_i \sigma_j$ and consider first the four terms where $i=j$. In this case, $\Omega=\sigma_i^2=\mathbb{I}$, and we simply end up with
\begin{equation} \label{sum1}
\sum_i q_i^2 \sigma_x = (q_0^2+q_x^2+q_y^2+q_z^2)\sigma_x.
\end{equation}
Then, we consider the six terms in the sum, where one of the indices is zero and the other is $\mu \in \{x,y,z\}$. In these cases, we have $\Omega=\sigma_\mu$. Recalling that $\sigma_\mu \sigma_x \sigma_\mu = s_x(\mu)\sigma_x$ with $s_x(\mu)=2\delta_{\mu,x}-1$, we obtain
\begin{equation} \label{sum2}
\sum_\mu q_0 q_\mu s_x(\mu) \sigma_x = 2q_0(q_x - q_y - q_z)\sigma_x.
\end{equation}
The remaining six terms with distinct non-zero indices are deduced noting that $\Omega=\sigma_\mu \sigma_\nu =\pm i\sigma_\alpha$ with $\alpha \neq \mu, \nu$. This leads to $\Omega\sigma_x\Omega=(\pm i\sigma_\alpha)\sigma_x(\pm i\sigma_\alpha)=-s_x(\alpha)\sigma_x$, and
\begin{equation} \label{sum3}
\sum_{\mu\neq\nu} -q_\mu q_\nu s_x(\alpha) \sigma_x = 2(q_xq_y  + q_xq_z - q_yq_z)\sigma_x.
\end{equation}
Finally, bringing the contribution of all the terms in the sum in Eq.~\eqref{super01} together using Eqs.~\eqref{sum1}-\eqref{sum3}, we identify the coefficient of the $\sigma_x\otimes \sigma_x$ term in Eq.~\eqref{evolvedrhosw} as
\begin{align}
    \kappa_x = {}& (q_0^2 + q_x^2 + q_y^2 + q_z^2) + 2q_0(q_x - q_y - q_z) \nonumber \\
               & + 2(q_x q_y + q_x q_z - q_y q_z),
\end{align}
which can be written, using Eq.~\eqref{alphap}, in the form
\begin{align}
    \kappa_x = {}& \big(1 - 2p(\alpha_y+\alpha_z)\big)^2 \nonumber \\
          &+ 4p^2\big(\alpha_x\alpha_y + \alpha_x\alpha_z - \alpha_y\alpha_z\big).   
\end{align}

A very similar calculation reveals that 
\begin{align}
    \kappa_y = {}& (q_0^2 + q_x^2 + q_y^2 + q_z^2) + 2q_0(-q_x + q_y - q_z) \nonumber \\
               & + 2(q_x q_y - q_x q_z + q_y q_z). \nonumber \\
               ={}& \big(1 - 2p(\alpha_x+\alpha_z)\big)^2 \nonumber \\
               &+ 4p^2\big(\alpha_x\alpha_y - \alpha_x\alpha_z + \alpha_y\alpha_z\big).
\end{align}

\section{Derivation of the state parameters \texorpdfstring{$\tau_x, \tau_y$ and $\tau_z$}{tau-x, tau-y and tau-z} under time-flipped Pauli channels} \label{app-b}

We can observe that the first term in the input state in Eq.~\eqref{Bellstate} remains the same under the time-flip $\mathcal{F}$ as a consequence of the unitality of Pauli channels. When it comes to the second term in Eq.~\eqref{evolvedrhotf}, the action of $\mathcal{F}$ on $\sigma_z \otimes \sigma_z$ is determined by the diagonal superoperators $\mathcal{F}_{00}$ and $\mathcal{F}_{11}$, due to the form of Eq.~\eqref{sup-mat2}. Specifically, 
\begin{align} \label{apndxb1}
\mathcal{F}(\sigma_z \otimes \sigma_z) ={}& |0\rangle\langle0| \otimes \mathcal{F}_{00}(\sigma_z) + |1\rangle\langle1| \otimes \mathcal{F}_{11}(-\sigma_z), \nonumber \\
={}& |0\rangle\langle0| \otimes \mathcal{F}_{00}(\sigma_z) - |1\rangle\langle1| \otimes \mathcal{F}_{11}(\sigma_z)
\end{align}
We first note that $\sigma_i=\sigma_i^\dagger$ and $\sigma_i^T=\sigma_i^*=\eta_i\sigma_i$, where $i \in \{0,x,y,z\}$ and $\eta=(1,1,-1,1)$. Then, looking at the definition of the superoperators $\mathcal{F}_{00}$ and $\mathcal{F}_{11}$ in Eq.~\eqref{sup-op-flip}, it can be easily seen that the Pauli channels are transposition invariant, i.e., $\Phi=\Phi^T$, and thus we have $\mathcal{F}_{00}=\mathcal{F}_{11}$. In accordance, it is possible to rewrite Eq.~\eqref{apndxb1} as
\begin{align}
\mathcal{F}(\sigma_z \otimes \sigma_z) ={}& \sigma_z\otimes\mathcal{F}_{00}(\sigma_z) = \sigma_z\otimes \Phi (\sigma_z) \nonumber \\
={}& \sigma_z\otimes \Lambda (\sigma_z) = \lambda_z(\sigma_z \otimes \sigma_z),
\end{align}
which shows that $\tau_z=\lambda_z$. Besides, the action of $\mathcal{F}$ on $\sigma_x \otimes \sigma_x$ and $\sigma_y \otimes \sigma_y$ are governed by the off-diagonal superoperators $\mathcal{F}_{01}$ and $\mathcal{F}_{10}$. To determine $\tau_x$, consider
\begin{equation}
\mathcal{F}(\sigma_x\otimes \sigma_x)= |0\rangle\langle1| \otimes \mathcal{F}_{01}(\sigma_x) + |1\rangle\langle0| \otimes \mathcal{F}_{10}(\sigma_x),
\end{equation}
which follows from the block off-diagonal form of $\sigma_x \otimes \sigma_x$. We also notice that the superoperators  $\mathcal{F}_{01}$ and $\mathcal{F}_{10}$ are indeed identical for Pauli channels, that is, $\mathcal{F}_{01}=\mathcal{F}_{10}$, thanks to the fact that $\sigma_i^T=\sigma_i^*=\eta_i\sigma_i$. Consequently,
\begin{equation} 
\mathcal{F}(\sigma_x\otimes \sigma_x)= \sigma_x \otimes \mathcal{F}_{01}(\sigma_x),
\end{equation}
which can be more explicitly written as
\begin{align} 
\mathcal{F}(\sigma_x\otimes \sigma_x) ={}& \sigma_x \otimes \sum_i q_i \eta_i (\sigma_i \sigma_x \sigma_i) \nonumber \\
={}&  \sigma_x \otimes \sum_i q_i \eta_i s_x(i) \sigma_x,
\end{align}
where $s_x(i)=2(\delta_{i,0}+\delta_{i,x})-1$, with $i \in \{0,x,y,z\}$. Then, summing over the four terms we identify that
\begin{equation} 
\tau_x=q_0+q_x+q_y-q_z=1-2q_z=1-2\alpha_z p.
\end{equation}
Lastly, using very similar arguments, we have
\begin{align} 
\mathcal{F}(\sigma_y\otimes \sigma_y) = \sigma_y \otimes \sum_i q_i \eta_i s_y(i) \sigma_y,
\end{align}
where $s_y(i)=2(\delta_{i,0}+\delta_{i,y})-1$, with $i \in \{0,x,y,z\}$. Then,
\begin{equation} 
\tau_y =q_0-q_x-q_y-q_z =1-2p,
\end{equation}
which completes the derivation of the coefficients.

\bibliography{bibliography}

\end{document}